\documentclass[10pt,twocolumn,twoside]{IEEEtran}
\usepackage{varwidth}
\usepackage{algorithm}
\usepackage{algpseudocode}
\usepackage{amsmath,epsfig,cite}
\usepackage{dsfont}
\usepackage{amsthm}
\usepackage{amssymb}
\usepackage{bm}
\usepackage{caption}
\usepackage{float}
\usepackage{graphics}
\usepackage{graphicx}
\usepackage{epsfig}
\usepackage{enumitem}
\usepackage{epstopdf}
\usepackage{latexsym}
\usepackage{mathtools}
\usepackage{pgfplots}
\usepackage{setspace}
\usepackage[utf8]{inputenc}
\newtheorem{definition}{Definition}

\DeclareMathOperator*{\argmax}{arg\,max}
\makeatletter
\newcommand*{\rom}[1]{\expandafter\@slowromancap\romannumeral #1@}
\makeatother

\begin{document}

\title{Multi-Player Multi-Armed Bandit Based Resource Allocation for D2D Communications}

\author{\IEEEauthorblockN{Anushree Neogi\IEEEauthorrefmark{1}, Prasanna Chaporkar\IEEEauthorrefmark{2} and
Abhay Karandikar\IEEEauthorrefmark{3}}

\IEEEauthorblockA{
\IEEEauthorrefmark{1}\IEEEauthorrefmark{2}\IEEEauthorrefmark{3}
Department of Electrical Engineering,
Indian Institute of Technology Bombay,
Mumbai -- 400076, India\\
Email:
(\IEEEauthorrefmark{1}anushreen,
\IEEEauthorrefmark{2}chaporkar,
\IEEEauthorrefmark{3}karandi)@ee.iitb.ac.in}}

\maketitle
\thispagestyle{plain}
\pagestyle{plain}

\begin{abstract}
Device-to-device (D2D) communications is expected to play a significant role in increasing the system capacity of the fifth generation (5G) wireless networks. To accomplish this, efficient power and resource allocation algorithms need to be devised for the D2D users. Since the D2D users are treated as secondary users, their interference to the cellular users (CUs) should not hamper the CU communications. Most of the prior works on D2D resource allocation assume full channel state information (CSI) at the base station (BS). However, the required channel gains for the D2D pairs may not be known. To acquire these in a fast fading channel requires extra power and control overhead. In this paper, we assume partial CSI and formulate the D2D power and resource allocation problem as a multi-armed bandit problem. We propose a power allocation scheme for the D2D users in which the BS allocates power to the D2D users if a certain signal-to-interference-plus-noise ratio (SINR) is maintained for the CUs. In a single player environment a D2D user selects a CU in every time slot by employing UCB1 algorithm. Since this resource allocation problem can also be considered as an adversarial bandit problem we have applied the exponential-weight algorithm for exploration and exploitation (Exp3) to solve it. In a multiple player environment, we extend UCB1 and Exp3 to multiple D2D users. We also propose two algorithms that are based on distributed learning algorithm with fairness (DLF) and $k$th-UCB1 algorithms in which the D2D users are ranked. Our simulation results show that our proposed algorithms are fair and achieve good performance.
\end{abstract}

\section{Introduction}
Device-to-device (D2D) communications is expected to increase the spectral efficiency of the fifth generation (5G) network. D2D users which are within a certain range can communicate with each other using the resources of the cellular users (CUs) in a Long Term Evolution (LTE) network. In such a network, the CUs are treated as primary users while the D2D users are treated as secondary users. In order to ensure that the quality of service (QoS) of the CUs do not get affected, efficient resource allocation algorithms need to be devised for the D2D users. Most of the previous research in D2D resource allocation assume the knowledge of all the channel gains. However, the channel gain between a D2D transmitter and its receiver and the channel gain between a CU and a D2D receiver is difficult to acquire because the cardinality of these channel gains is high. It is the number of CUs times the number of D2D receivers plus the number of D2D pairs. To convey these to the BS requires extra power and control overhead. We consider a realistic situation in which these gains are unknown at the BS, which is referred to as partial channel state information (CSI). We solve the power and resource allocation problem for the D2D users with partial CSI by formulating it as a multi-player multi-armed bandit (MP-MAB) problem.

\subsection{Motivation}
\indent Most of the applications of MP-MAB in  wireless communications are in the field of cognitive radio networks (CRNs). In a CRN, the mean of a channel (arm) does not vary from one secondary user to the other. Thus, the optimal channel is the same for all of them. If a perfect collision model is assumed \cite{a}, then if they collide the reward obtained by them is zero. Since UCB1 will determine the best arm among all the arms, if it is applied to this problem, the secondary users will collide with each other as they contend for the same optimal arm. This has motivated researchers to propose  variants of UCB1 like distributed learning algorithm with prioritization (DLP), distributed learning algorithm with fairness (DLF) \cite{a} and $k$th-UCB1 \cite{b}. In these algorithms, the secondary users have different ranks such that a secondary user which has a rank of $k$ will select the arm with the $k^{th}$ largest mean reward  in the long run instead of the arm that gives it the highest mean reward. Thus, the chances of secondary users to collide with each other decreases. DLP is extended to DLF so that fairness is ensured among the secondary users. This is done by changing the rank of each user in a round robin (RR) manner such that they have unique ranks in each time slot. \\
\indent The D2D resource allocation problem is significantly different from the channel selection problem considered in a CRN. In this paper, we consider multiple D2D users and CUs in a macro cell. We utilize the MP-MAB framework to solve the problem of power and resource allocation for the D2D users. We assume that a CU is allocated only one resource block. We also assume that a D2D user can be allocated the resource block of only one CU. Our objective is to maximize the expected value of the cumulative sum throughput of the D2D users upto a time horizon.  For power allocation to a D2D user the signal-to-noise ratio (SNR) of a CU, whose resource block is allocated to the D2D user, should be greater than a certain threshold. Then only power can be allocated to the D2D user so as to ensure that the CU's signal-to-interference-plus-noise ratio (SINR) is this minimum threshold. Thus, we guarantee a minimum SINR to the CUs whose resource blocks are allocated to the D2D users. We model the instantaneous reward received by a D2D user when it selects a CU's  resource block as its throughput, normalized to lie in [0, 1]. Therefore, the mean reward of an arm differs from one D2D user to the other. If the number of CUs are more than the D2D users then the chances of each D2D user's optimal arm to be a different CU is more. Hence, it is appropriate to employ the UCB1 algorithm to this problem.
 
\subsection{Related Work}
\indent In \cite{c}, the authors have proposed two index based distributed learning algorithms $\rho^{PRE}$ and $\rho^{RAND}$ for channel selection in a CRN. $\rho^{PRE}$ is based on the $\epsilon_n$-greedy algorithm of \cite{d} where the secondary users are ranked a-priori. The $\rho^{RAND}$ algorithm is based on adaptive randomization in which every user randomizes its channel selection if there is a collision in the previous time slot. In \cite{e}, another index based algorithm called dUCB4 is investigated which order log-squared regret growth with a certain time horizon. However, the players are allowed to communicate with each other which increases the communication overhead. In \cite{a}, the authors present a generalization of UCB1 called selective learning of the $K^{th}$ largest expected rewards (SL(K)) where the secondary user learns to select the arm which gives the $K^{th}$ largest expected  reward among all the arms. Moreover, they extend this algorithm to multiple users in which the users are ranked and propose a policy called DLP. However, to ensure fairness among the users they propose DLF which rotates the rank among the users. They have proved that DLF is order-optimal.\\
\indent Another order optimal policy is considered in \cite{f} known as time division fair sharing (TDFS) which guarantees fairness among the secondary users. The users have different offsets in their time division selection schedule to avoid collisions during channel selection. However, it is computationally complex. In \cite{g}, an algorithm based on deterministic sequencing of exploration and exploitation (DSEE) is developed in which time is divided into exploitation and exploration sequences. The main design criterion is to determine the cardinality of the exploration sequence. However,  knowledge of a lower bound on the difference in the mean rewards of the best and the second best arms is required which is difficult to obtain a-priori. In \cite{b},a secondary user is assigned a rank $k$ a-priori and an algorithm called $k$th-UCB1 is proposed.\\
\indent The adversarial MAB problem is different from the stochastic MAB problem in that there is no assumption on the distribution of the reward process of each arm. When the rewards of all the arms can be observed by a player (full information game), the Hedge algorithm \cite{h} can be used. A player chooses an action according to a probability distribution over the arms. It is a modification of the weighted majority algorithm \cite{i}. The Exponential-weight algorithm for Exploration and Exploitation (Exp3) \cite{j} in turn is a modification of the Hedge algorithm. These algorithms are weighted average prediction (WAP) algorithms in which the probability of choosing an arm is computed by assigning weights to each arm from which a probability distribution over the arms is calculated. \\
\indent Though MAB has been applied to CRNs, its application to other wireless communication problems is limited. We next discuss a few works pertaining to the application of MAB to 4G/5G networks. In \cite{k}, the authors have used Exp3.M algorithm to address the problem of efficiently activating or deactivating small cells in a macrocell dynamically by modeling it within the framework of a combinatorial MP-MAB. In \cite{l}, the relay selection problem is modeled as a stochastic covariate MAB where side information is available to the players before each trial. In \cite{m}, the authors have considered the problem of selecting sub-bands for each picocell in an LTE network. They have solved this problem using UCB1 algorithm. To allocate the resources of the sub-band chosen by a picocell to its users, each picocell utilizes the proportional fair scheduling algorithm. Among the recent works in the field of D2D communications, \cite{n} addresses the problem of selecting the CU transmission mode or the D2D transmission mode by modeling it as a two-armed Levy-bandit game. The authors have considered the number of interferers to be random. They have considered two cases of the game model, multiple independent players and multiple cooperative players. The optimal strategy in both the cases is a cut-off strategy for all the users. In \cite{o}, the authors have considered resource allocation for the D2D users by formulating it within the framework of an MP-MAB game with side information. They have combined no-regret learning with calibrated forecasting and proved that the empirical joint frequencies of the game converge to a set of correlated equilibria. However, even this algorithm is computationally expensive.

\subsection{Contributions}
The main contributions of this paper are as follows.
\begin{itemize}[leftmargin=*]
\item Ours is the first work which models the D2D  power and resource allocation problem within an MP-MAB framework. Our proposed algorithms ensure that the D2D users reuse the resources of the CUs efficiently without hindering CU communications. Our work differs from \cite{o} as the authors have not considered the reuse of CUs' resources. They assume that the D2D users use only the vacant channels of the CUs. 
\item   We propose a power allocation algorithm for the D2D users that ensures a minimum quality of service (QoS) to a CU so that its communications are not hampered due to interference from a D2D user.
\item Our work differs from \cite{m} as follows. In \cite{m}, the UCB1 algorithm is used but when multiple picocells select the same sub-band and the resources of this sub-band are allocated to users, those users which are allocated the same resource block suffer from inter-picocell interference when they transmit. However, in our case we employ the perfect collision model of \cite{a} in which if multiple D2D users select the same CU's resource block, i.e. collisions occur, then the D2D users do not transmit. This ensures that there is no inter-D2D interference. Moreover, the CU also does not get affected due to interference from multiple D2D users. 
\item We apply DLF and $k$th-UCB1  which we extend to multiple D2D users and modify to ensure fairness as in \cite{a}.
Since the D2D resource allocation problem can also be solved within the framework of an adversarial bandit problem we have applied the Exp3 algorithm for allocating resources to the D2D users.
\end{itemize} 
\indent This paper is organized as follows. In Section II, we discuss the system model of an underlay D2D network and the problem formulation in an MP-MAB setting. In Section III, we propose two power and resource allocation algorithms for a single D2D user based on UCB1 and Exp3. In Section IV, we propose four  power and resource allocation algorithms for multiple D2D users by extending UCB1 and Exp3 to multiple  users and applying DLF and $k$th-UCB1 to the multi-player setting. We illustrate the performance of our proposed algorithms through simulation results in Section V and conclude the paper in Section VI. 
 	\begin{figure}[t]
 	\centering
 	\scalebox{0.8}{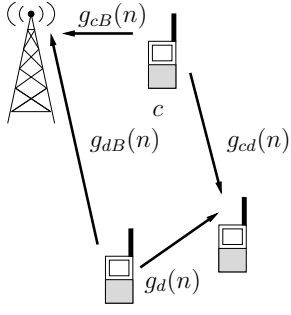}
 	\caption{Resource reuse between a CU and a D2D user.}
 	\label{fig:1}
 	\end{figure}	
    
\section{System Model and Problem Formulation}
\subsection{Network Model}
We consider $N_C$ CUs and $N_D$ D2D users in a macrocell comprising of the sets $\mathcal{N}_C$ and $\mathcal{N}_D$ respectively. We assume that the CUs are allocated uplink resource blocks by the base station (BS) in every subframe. In an LTE network, time is divided into subframes which we denote by $n$. We consider path loss, shadowing and fast fading. The channel gain between a CU $c$ and the BS $B$ is represented by $g_{cB}(n)$ while that between a D2D user's transmitter and receiver by $g_{d}(n)$. The channel gain of the interfering link between a CU $c$ and a D2D user $d$'s receiver is given by 
$g_{cd}(n)$ and that between the D2D user's transmitter and the BS is given by $g_{dB}(n)$, which is shown in Fig. 1. 
In an LTE network, a user equipment (UE) sends its channel quality information (CQI) to the BS either periodically or aperiodically. Thus, the channel gains $g_{cB}(n)$ and $g_{dB}(n)$ are known at the BS. However, it is difficult to obtain the channel gains $g_{d}(n)$ and $g_{cd}(n)$. A CU's transmit power $P_c$ is assumed to be constant. We assume that each CU is allocated one resource block in each subframe and each D2D player can be allocated only one resource block.

\subsection{MP-MAB Framework}
 We formulate the D2D resource allocation problem within the framework of an MP-MAB problem where the players are the D2D users.  The arms of the bandit are the available CUs. The action $a_d(t)$ of a D2D player $d$  refers to the selection of a CU.  When a D2D player $d$ selects an arm (a CU $c$) it obtains a random reward $X_{d,c}(n)$, with mean $\mu_{d,c}=\mathbb{E}[X_{d,c}(n)]$, from an unknown stationary distribution of the reward process of the arm. We model the reward $X_{d,c}(n)$  of a D2D player $d$ when it selects a CU $c$ to be its throughput $r_d(n)$ normalized to lie within the range of $[0,1]$. The normalized throughput is denoted by $\tilde{r}(n)$. When collisions occur we assume that the D2D players which select the same CU receive a reward of zero. The power allocated to each D2D player depends on whether a minimum SINR can be guaranteed to a CU. The power allocated to the D2D player therefore depends on the channel gains $g_{cB}(n)$, $g_{dB}(n)$ and the power transmitted by the CU $P_c$. Thus, the throughput and therefore the reward depends on the channel gains $g_{cB}(n)$, $g_{d}(n)$, $g_{cd}(n)$, $g_{dB}(n)$ and the power transmitted by
 the CU.  Therefore, these decide the reward process corresponding to each arm, that is, a CU. Thus, the distribution of the reward process of each arm is different for different D2D players. Since the channel gains are independent and identically distributed (IID) random variables, the reward process of an arm is also IID for each D2D player.  \\  
\indent A local policy $\pi_d=\{\pi_d(n)\}_{n\geq 1}$ of a D2D player $d$ is a sequence of functions where $\pi_d(n)$ maps the player's previously observed rewards and actions to the present action $a_d(n)$ of the player at time instant $n$. It is a decision rule that specifies for each D2D player which action to take at each time instant.
A policy $\pi$ is a therefore a concatenation of the local policies of all the players $\pi = [\pi_1, \dots, \pi_{N_D}]$. 

  \begin{definition}
  (\cite{d}) The optimal arm for a D2D player   $d$ is the CU $c_d^*$ whose expected reward is the highest among all the arms (CUs). Thus, the optimal CU is given by,
  \begin{align}
  c_d^*=\argmax_{c \in \mathcal{N}_C} \: \mu_{d,c}.
  \end{align}
  \end{definition}

  \begin{definition}
  For a given policy $\pi$, we define its regret $R_{\pi}(T)$ as the difference in the the expected total reward of the D2D players upto time $T$ that can be obtained by them when they select their optimal arms and the expected total reward of the D2D players obtained under policy $\pi$ upto time $T$. It is given by,
  \begin{align} \label{eqn:2}
R_{\pi}(T)=T\sum\limits_{d =1}^{N_D} \mu_{d,c_d^*} - \mathbb{E}_{\pi}\bigg[ \sum \limits_{n=1}^{T}\sum\limits_{d =1}^{N_D}X_{d,\pi_d(n)}(n)\bigg],
  \end{align}
  where $\mu_{d,c_d^*}$ is the mean of the optimal arm $c_d^*$ for the D2D player $d$.
  \end{definition}
\noindent  The performance of a policy $\pi$ is measured by the regret $R_{\pi}(T)$. By observing how the regret changes with $n$ for different policies we can compare different regret graphs.\\
\indent If the D2D players are ranked in each subframe, then the definition of regret changes as per \cite{a}.
\label{eqn:2}
\begin{definition}
(\cite{a}) If the rank of a D2D player $d$ is $K$ in a subframe, then let the arm which has the $K$-th largest expected reward be $c_K^*$ and the mean of this arm be $\mu_{d,{c_K^*}}$. Then the regret is given by,
\begin{align} 
R_{\pi}(T) =\sum\limits_{d=1}^{N_D} \sum\limits_{n=1}^{T} \Bigl|\mu_{d,c_K^*} - \mathbb{E}_{\pi} \sum\limits_{c=1}^{N_C} X_{d,c}(n)\mathds{1}_{d,c}(n)\Bigr|, 
\end{align}
\end{definition}
\noindent where $\mathds{1}_{d,c}(n)$ is an indicator random variable which is 1 if D2D player $d$ is the only player that selects the CU $c$ as per the policy $\pi$ and zero otherwise.\\
\indent For an adversarial bandit, there are no statistical assumptions in the way the rewards are generated  \cite{j}. The regret definition differs from Eqns. 5.2 and 5.3. The optimal CU's definition also changes as follows. 
  \begin{algorithm}[t]
  \caption{Power and resource allocation for a D2D Player $d$ in each subframe with UCB1 algorithm}\label{alg:1}
  \begin{algorithmic}[1]
  \State Initialize: $y_{d,c}(n)\gets0$, $\hat{\mu}_{d,c}(n)\gets0,\qquad  \forall \: c \: \in \mathcal{N}_C$
  \If{$1 \le\: n \le \: N_C$}
  \State \parbox[t]{\dimexpr\linewidth-\algorithmicindent}{The D2D player $d$ chooses CU $c=n$.\strut}
  \Else
  \State \parbox[t]{\dimexpr\linewidth-\algorithmicindent}{It calculates the UCB1 index as per Eqn. 10.\strut}
  \State \parbox[t]{\dimexpr\linewidth-\algorithmicindent}{It chooses a CU $c$ whose UCB1 index is maximum. \strut}
  \EndIf
  \State It communicates its choice of $c$ to the BS.
  \State \parbox[t]{\dimexpr\linewidth-\algorithmicindent}{The BS allocates power, \par $P_d(n)\gets$ \textsc{PowerAlloc}$(c,g_{cB},g_{dB})$.\strut}
  \State The D2D player observes a throughput of $r_d(n)$.
  \State It normalizes its throughput $r_{d}(n)$ to get its reward $X_{d,c}(n)=\tilde{r}_d(n)$ .
  \State \parbox[t]{\dimexpr\linewidth-\algorithmicindent}{It increments its counter $y_{d,c}(n)$.\strut}
  \State \parbox[t]{\dimexpr\linewidth-\algorithmicindent}{It calculates the empirical mean reward $\hat{\mu}_{d,c}(n)$ of the arm $c$ as per Eqn. 9.\strut}
  \end{algorithmic}
  \end{algorithm}
\begin{definition}
(\cite{j}) For an adversarial bandit framework, the optimal CU $c^*$ for a D2D player is one which maximizes its return, i.e. the sum of rewards it receives upto time $T$.
\end{definition}
\begin{definition}
(\cite{h}) In an adversarial bandit framework, let the actions taken by a D2D player till time $T$ be $a(1),a(2),...,a(T)$, then the expectation of the total reward obtained with the policy $\pi$ upto time $T$ is,
\begin{align}
\mathbb{E}[G_\pi(t)]=\mathbb{E}_{a(1),a(2),...,a(T)}\bigg[\sum_{n=1}^{T}X_{a(n)}(n)\bigg],\nonumber
\end{align}
where $X_{a(n)}(n)$ is the reward received by the D2D player when it takes an action $a(n)$.
Let the expected total reward from the optimal arm till time $T$ be,
\begin{align}
G^*(T)=\max\limits_{c \in \mathcal{N}_C}\: \mathbb{E}_{a(1),a(2),...,a(T)}\bigg[\sum_{n=1}^{T}X_{c}(n)\bigg].\nonumber
\end{align}
The regret for a D2D player $d$ is given by,
\begin{align}
R^d_{\pi}(T)=G^*(T)-\mathbb{E}[G_\pi(T)].\nonumber
\end{align}
The regret for multiple D2D players is the sum of individual regrets of each D2D player and is given by,
\begin{align}
R_{\pi}(T)=\sum\limits_{d=1}^{N_D}R^d_{\pi}(T).
\end{align}
\end{definition}
Minimizing the regret of Eqn. 2, 3 or 4 is equivalent to maximizing the expected total reward of the D2D players upto time $T$.
\noindent Our objective is to determine a policy which maximizes the expected value of the cumulative sum throughput of the D2D players till time $T$.  \\
\indent We next propose two power and resource allocation policies for a single D2D player and then extend them to multiple D2D players.

\section{Allocation Policies for a Single D2D Player}
The D2D resource allocation problem can be solved within the framework of both a stochastic MAB (UCB1 algorithm \cite{d}) and an adversarial MAB (Exp3 algorithm \cite{j}). We propose two power and resource allocation policies based on the UCB1 and Exp3 algorithms. In the UCB1 based algorithm the D2D player selects a CU as per the UCB1 index. It selects that CU in each subframe which has the maximum UCB1 index.  In Exp3 each D2D player selects a CU randomly as per a probability distribution over the CUs. We next discuss these two policies. 

\subsection{Power and Resource Allocation Based on UCB1}
\subsubsection{Initialization Phase}
 The UCB1 algorithm starts with an initialization phase (refer \emph{Algorithm 1}) in which a D2D player $d$ samples all the arms (the CUs) sequentially in the first $N_C$ subframes. In every subframe $n$, it conveys its selection of a CU $c$ to the BS. The BS then allocates power to it as follows.
  \begin{algorithm}[t]
  \caption{Power allocation}\label{alg:2}
  \begin{algorithmic}[1]
  \Statex \textbf{Input:  }$c,g_{cB},g_{dB}$
  \Statex \textbf{Output:  }$P_d(n)$
  \Function{PowerAlloc}{$c,g_{cB},g_{dB}$}
   \If{D2D player $d$ suffers from collision}
  \State $P_{d}(n)\gets 0$
  \Else
  \State $P_{d}(n) \gets \bigg(\frac{P_{c}g_{cB}(n)}{{\gamma}^{tgt} g_{dB}(n)} - \frac{{\sigma}^{2}_{BS}}{g_{dB}(n)}\bigg)^+$
  \If{$P_{d}(n)>P_{max}$}
  \State $P_{d}(n)\gets P_{max}$
  \EndIf
  \EndIf 
  \EndFunction
  \end{algorithmic}
  \end{algorithm}
 \subsubsection{Power Allocation  (\cite{p})}
 When a D2D user $d$ selects a CU $c$'s resource block, the SINR of the CU $\gamma_{c}(n)$ must be greater than equal to the SINR threshold $\gamma^{tgt}$, 
\begin{align}\label{eqn:3}
        \gamma_{c}(n)=\frac{P_{c}g_{cB}(n)}{\sigma^2_{BS} + P_{d}(n)g_{dB}(n)} \ge {\gamma}^{tgt}, 
\end{align}
where $\sigma^2_{BS}$ is the average noise power at the BS. Then,
\begin{align}\label{eq:4}
P_{d}(n) \le \bigg( \frac{P_{c}g_{cB}(n)}{{\gamma}^{tgt} g_{dB}(n)} - \frac{\sigma^{2}_{BS}}{g_{dB}(n)}\bigg).
\end{align} 
So, the maximum power that can be allocated 
to the D2D player is,   
\begin{align}\label{eq:4}
P_{d}(n) = \bigg( \frac{P_{c}g_{cB}(n)}{{\gamma}^{tgt} g_{dB}(n)} - \frac{\sigma^{2}_{BS}}{g_{dB}(n)}\bigg)^+.
\end{align}
This means that $\gamma_c$ is equal to $\gamma^{tgt}$. Note that this also implies that the SNR of a CU (without interference from a D2D player) should be greater than $\gamma^{tgt}$, then only power can be allocated to the D2D player.  This condition is checked by the BS. If the SNR is less than $\gamma^{tgt}$, then $P_d(n)$ will become negative and if it is equal to $\gamma^{tgt}$, then $P_d(n)$ will become zero in order to ensure $\gamma_c=\gamma^{tgt}$. This implies that in both these cases power cannot be allocated to the D2D player. If $P_d(n)$ exceeds the maximum allowed transmission power of a UE  $P_{max}$, then $P_d(n)$ is limited to $P_{max}$. This means that $P_d(n)$ calculated from Eq. 5.7 can be large but in practice a lesser value of $P_{max}$ has to be allocated which causes $\gamma_c$ to become more than $\gamma_{tgt}$  (refer \emph{Algorithm 2}). \\
\indent The throughput observed by a D2D player is given by,
\begin{equation}\label{eq:5}
r_d(n)=B\log_2\bigg(1+\frac{P_{d}(n)g_{d}(n)}{{\sigma}^{2}_{D2D}+P_{c}g_{cd}(n)} \bigg),
\end{equation}
where $B$ is the transmission bandwidth and $\sigma^{2}_{D2D}$ is the average noise power at the D2D receiver. \\

\subsubsection{Resource Allocation}
We model the reward $X_{d,c}(n)$ to be the normalized throughput $\tilde{r}_d(n)$ which the D2D player $d$ receives when it selects CU $c$. $\tilde{r}_d(n)$ is the throughput $r_d(n)$ normalized to lie between 0 and 1. The reward  $X_{d,c}(n)$ is therefore a random variable drawn from the distribution of the reward process of an arm $c$.  \\
\indent After the initialization phase of $N_C$ subframes, the D2D player $d$ stores one sample of the reward from each of the $N_C$ arms. Thereafter in every subframe $n$, the D2D player averages its previous and present rewards accrued by it whenever it selects a CU $c$ till subframe $n$. This is nothing but the empirical mean reward $\hat{\mu}_{d,c}(n)$ which is calculated as per the following Monte Carlo equation,
  \begin{align}
 \hat{\mu}_{d,c}(n)= \bigg(1-\frac{1}{y_{d,c}(n-1)}\bigg)\hat{\mu}_{d,c}(n-1)+\frac{1}{y_{d,c}(n-1)}\tilde{r}_d(n),
  \end{align}\label{eqn:9}
\hspace{-2mm}where $y_{d,c}(n-1)$ is the number of times that it selects a CU $c$ till the previous subframe $n-1$. The UCB1 index is a function of the previous empirical mean reward $\hat{\mu}_{d,c}(n-1)$ and $y_{d,c}(n-1)$ corresponding to a CU $c$. A D2D player calculates the UCB1 index of each CU which is given by,
 \begin{align}
 \hat{\mu}_{d,c}(n-1) + \sqrt{\frac{2 \ ln(n)}{y_{d,c}(n-1)}}.
 \label{eqn:6}
 \end{align}
The policy chooses that CU in each subframe $n$ whose UCB1 index is the maximum among the UCB1 indices of all the CUs. This power and resource allocation algorithm is optimal as it achieves logarithmic regret \cite{d}, $O (\,\text{ln}\, T)$.

\subsection{Power and Resource Allocation Based on Exp3}
We now solve the D2D power and resource allocation problem for a single D2D player by applying the Exp3 algorithm (refer \emph{Algorithm 3}). The arms of the adversarial bandit are the CUs. We consider the adversary to be the channel which decides the rewards to be given to the D2D player when it selects a CU. We model the reward $X_{d,c}(n)$ obtained by a D2D player $d$ from each arm $c$ as a Bernoulli random variable whose distribution is unknown. When the D2D player achieves a throughput of $r_d(n)$ greater than equal to the guaranteed throughput $r'$, its reward is one else it is zero. If $X_{d,c}(n)$ is one most of the time when it selects a CU $c$ as compared to the other CUs, it implies that this choice of CU is better than the others because on an average it gets a better throughput with this CU. This is equivalent to maximizing the expected cumulative reward of the D2D player till a certain time. \\
\indent The Exp3 algorithm selects a CU as per the probability density function (PDF), $p_d(n)=\{p_{d,1}(n), \cdots, p_{d,c}(n), \cdots, p_{d,N_C}(n) \}$ over the CUs. This PDF is computed by the D2D player by assigning a weight $w_{d,c}(n)$ to each CU $c$. The weights of all the CUs are initialized to one, that is, $w_{d,c}(n)=1$. The PDF $p_d(n)$ is computed from the weights of the CUs as,
\begin{eqnarray}
p_{d,c}(n)=(1-\alpha)\frac{w_{d,c}(n)}{\sum_{c=1}^{N_C}w_{d,c}(n)}+\frac{\alpha}{N_C},
\end{eqnarray}
where $\alpha \in (0,1]$. The factor $\alpha/N_C$ ascertains that the D2D player explores all the arms often enough. 
\begin{algorithm}[t]
\caption{Power and resource allocation for a D2D Player $d$ with Exp3 algorithm in each subframe}\label{alg:1}
\begin{algorithmic}[1]
\State Initialize: $\alpha \in (0,1]$, $w_{d,c}(n)\gets1,  \:  \forall \: c \: \in \mathcal{N}_C$
\State The D2D player computes $p_d(n)$ from the weights of the CUs as per Eqn. 11.
\State It chooses a CU $c$ randomly as per the PDF $p_d(n)$.
\State It conveys its choice of CU $c$ to the BS.
\State \parbox[t]{\dimexpr\linewidth-\algorithmicindent}{The BS allocates power, \par $P_d(n)\gets$ \textsc{PowerAlloc}$(c,g_{cB},g_{dB})$.\strut}
\State The D2D player observes a throughput of $r_d(n)$.
\If{$r_d(n) \ge  r'$ }
\State $X_{d,c}(n)\gets1$ 
\Else
\State $X_{d,c}(n)\gets0$ 
\EndIf
\State For all the other CUs, $X_{d,c}(n)/p_{d,c}(n)=0$ 
\State The D2D player updates the weights $w_{d,c}(n)$ of all the CUs as per Eqn. 12.
\end{algorithmic}
\end{algorithm}
This ensures that there are sufficient samples of rewards from each arm to determine $p_d(n)$. The parameter $\alpha$ decides the trade-off between exploration and exploitation. The D2D player selects a CU $c$ according to the PDF $p_d(n)$ and conveys it to the BS. The BS checks if an SINR of $\gamma^{tgt}$ can be guaranteed to the CU. If  so, then the BS allocates a power of $P_d(n)$ as per the power allocation algorithm discussed for the UCB1 based policy (refer \emph{Algorithm 2}). The D2D player then transmits using CU $c$'s resource block and obtains a throughput of $r_d(n)$ which is given by Eqn. 8. The reward $X_{d,c}(n)$ received by the D2D player is one if its throughput is greater than equal to a target rate of $r'$, else its reward is zero. This D2D player then updates the weight $w_{d,c}(n)$ assigned to this CU to $w_{d,c}(n+1)$  by multiplying $w_{d,c}(n)$ with an exponent which is a function of the reward $X_{d,c}(n)$ and $p_{d,c}(n)$ the probability of selecting this CU,
\begin{eqnarray}
w_{d,c}(n+1)=w_{d,c}(n)exp\bigg(\frac{\alpha X_{d,c}(n)}{ N_Cp_{d,c}(n)}\bigg).
\end{eqnarray}

\section{Allocation Policies for Multiple D2D Players}
We next address the D2D resource allocation problem for multiple D2D players. First, we discuss the index based algorithms. We extend UCB1 from a single player to a multi-player setting and refer it as multi-player UCB1 (MP-UCB1). Next, we employ the DLF and $k$th-UCB1 algorithms  for resource allocation to multiple D2D players. We extend the $k$th-UCB1 algorithm for a single player given in \cite{b} to multiple players and refer to it as fair $k$th-UCB1. We also extend the Exp3 algorithm from a single player to a multi-player setting and solve the D2D resource allocation problem within an adversarial bandit framework in which the channel and the other players together act as an adversary.
 \begin{algorithm}[t]
  \caption{Power and resource allocation for a D2D Player $d$ in each subframe in a multi-player setting }\label{alg:1}
  \begin{algorithmic}[1]
   \State Initialize: $y_{d,c}(n)\gets0$, $\hat{\mu}_{d,c}(n)\gets0$, $\: \forall \: c \: \in \mathcal{N}_C$
  \If{$1\:  \le n \le \: N_C$}
  \State \parbox[t]{\dimexpr\linewidth-\algorithmicindent}{D2D player $d$ chooses $c\gets ((n+d) \, mod \, N_C)+1$\strut}
  \Else
  \State \parbox[t]{\dimexpr\linewidth-\algorithmicindent}{It calculates the UCB1 index as per Eqn. 10.\strut}
  \State \parbox[t]{\dimexpr\linewidth-\algorithmicindent}{It chooses a CU $c$ as per the index based algorithm.\strut}
  \EndIf
  \State It communicates its choice of $c$ to the BS.
  \State \parbox[t]{\dimexpr\linewidth-\algorithmicindent}{The BS allocates power, \par $P_d(n)\gets$ \textsc{PowerAlloc}$(c,g_{cB},g_{dB})$.\strut}
  \State The D2D player observes a throughput of $r_d(n)$.
  \State It normalizes its throughput $r_d(n)$ to get the reward $X_{d,c}(n)=\tilde{r}_{d}(n)$.
\State \parbox[t]{\dimexpr\linewidth-\algorithmicindent}{It increments its counter $y_{d,c}(n)$.\strut}
\State \parbox[t]{\dimexpr\linewidth-\algorithmicindent}{It calculates the empirical mean reward $\hat{\mu}_{d,c}(n)$ as per Eqn. 9.\strut}
  \end{algorithmic}
  \end{algorithm}
\subsection{Index Based Allocation Policies}
We initialize MP-UCB1 and $k$th-UCB1 for multiple players in the same way as DLF. Thus, the initialization phase is the same for all the three policies (refer \emph{Algorithm 4}). The power allocation algorithm is also the same for them. They differ in the way the CUs' resources are allocated to the D2D players over the subframes and that the players are ranked in  DLF and fair $k$th-UCB1 algorithms unlike MP-UCB1.\\

\subsubsection{Initialization Phase}
In the first $N_C$ subframes, every D2D player selects the CUs sequentially such that they sample one value of the reward from each of the CUs. In a subframe $n$, the D2D players choose the CUs in a RR order so that they don't collide with each other, $c= ((n+d) \, mod \, N_C)+1$. After this initialization phase, each of them conveys its selection of CU to the BS. The BS then allocates power to them. \\

\subsubsection{Power Allocation}
The power allocation algorithm is the same as for the single player setting. If a D2D player $d$ selects a CU $c$, it is allocated a power of $P_d(n)$ as per Eqn. 7 if a minimum SINR can be guaranteed to the CU. If $P_d(n)$ exceeds the maximum allowed transmission power of a UE $P_{max}$, then it is limited by $P_{max}$ (refer \emph{Algorithm 2}). A D2D player transmits at a rate of $r_d(n)$. Its normalized rate $\tilde{r}_d(n)$ is its reward. If multiple D2D players collide, then they get a reward of zero (perfect collision model \cite{a}). This implies that the transmitted powers of the D2D players that collide become zero.  \\
  
\subsubsection{Resource Allocation}
 After $N_C$ subframes, each D2D player chooses its CU as per the index of MP-UCB1, DLF and fair $k$th-UCB1 algorithms. Every time a D2D player chooses a CU $c$, it averages its rewards obtained with this CU till subframe $n$ as per Eqn. 9 and computes the empirical mean reward $\hat{\mu}_{d,c}(n)$. It also updates $y_{d,c}(n)$, the number of times that it selects the CU $c$ till subframe $n$. The index of each algorithm depends on these two quantities based on which a D2D player selects its CU.\\\\
 \noindent \emph{1. MP-UCB1 Algorithm:} 
Each D2D player runs UCB1 at its end.
A D2D player $d$ selects that CU $c$ which has the maximum UCB1 index .\\\\
\noindent \emph{2. DLF Algorithm:} Unlike MP-UCB1, in this algorithm a D2D player $d$ is assigned a rank  $K=(n+d)\: mod \:N_D+1$ in subframe $n$ that changes over subframes in a RR order. In a subframe $n$, every D2D player is assigned a unique rank. A D2D player 
 ranked $K$ in subframe $n$ tries to select a CU that gives it the $K^{th}$ largest mean reward. It sorts the UCB1 indices of all the CUs and selects the first $K$ CUs with the $K$ largest UCB1 indices to form a set $\mathcal{O}_K$. It selects a CU $c$ from $\mathcal{O}_K$ which minimizes the following index,
\begin{align}
 \hat{\mu}_{d,c}(n-1) - \sqrt{\frac{2 \ ln(n)}{y_{d,c}(n-1)}}.\label{eqn:8}
\end{align}
\\
\noindent \emph{3. Fair $k$th-UCB1 Algorithm:} In this algorithm also, each D2D player is ranked and its rank $K=(n+d)\: mod \:N_D+1$ changes as per RR in every subframe. A D2D player that is ranked $K$ in subframe $n$, chooses the $K$ CUs with the $K$ largest UCB1 indices to form a set $\mathcal{O}_K$. Then, it selects the CU $c'$ in the set $\mathcal{O}_K$ which has the minimum UCB1 index.
Now, with a probability of $1-\epsilon_n$, it chooses $c'$ else it selects a CU uniformly at random from $\mathcal{O}_K$ with a probability of $\epsilon_n$ where $\epsilon_n=min(\beta/n,1)$ . If  $\epsilon_n$ is a constant,  then it results in a linear regret \cite{d}. Therefore, $\epsilon_n$ is set such that it decreases with time $n$. This ensures a logarithmic regret.

\subsection{Adversarial Bandit Based Allocation Policy}
We now extend the Exp3 based policy discussed for a single player to multiple players  (refer \emph{Algorithm 3}). A D2D player selects a CU as per the PDF $p_d(n)$ over the CUs as discussed for the single player policy. In a multi-player setting not only does the channel decide the rewards, other players also decide the rewards that a D2D player obtains. When the D2D players convey their choice of CUs to the BS, the BS allocates power as per the power allocation algorithm discussed for the single player policy (refer \emph{Algorithm 2}). 
 First, it checks for collisions among the players. If they collide, their allocated power is zero. If the D2D player suffers no collision, then the BS checks  if the CU it has selected can be guaranteed an SINR of $\gamma^{tgt}$. If this can be guaranteed,  the D2D player is allocated power. If not, it is not allocated any power. Thus, its transmission rate $r_d(n)$ becomes zero. If the allocated power $P_d(n)$ is greater than $P_{max}$, then $P_d(n)=P_{max}$. If $r_d(n)$ is greater than $r'$, the D2D player gets a reward of one, else it gets a reward of zero.
                            
\section{Results}
 We consider a macrocell of radius 250 m with $N_C$ = 20 CUs and $N_D$ = 5 D2D players uniformly distributed in it. The D2D receiver is distributed uniformly around its D2D transmitter within a range of 50 m. As per LTE specifications, the path loss model is $PL=128.1+37.6 \ log\ (d)$ \cite{q}. We model shadow fading with a lognormal random variable whose standard deviation is 8 dB. We assume that the channel is a fast fading channel. 
                     \begin{figure}[t]
                    \begin{center}
                  \includegraphics[scale=0.56]{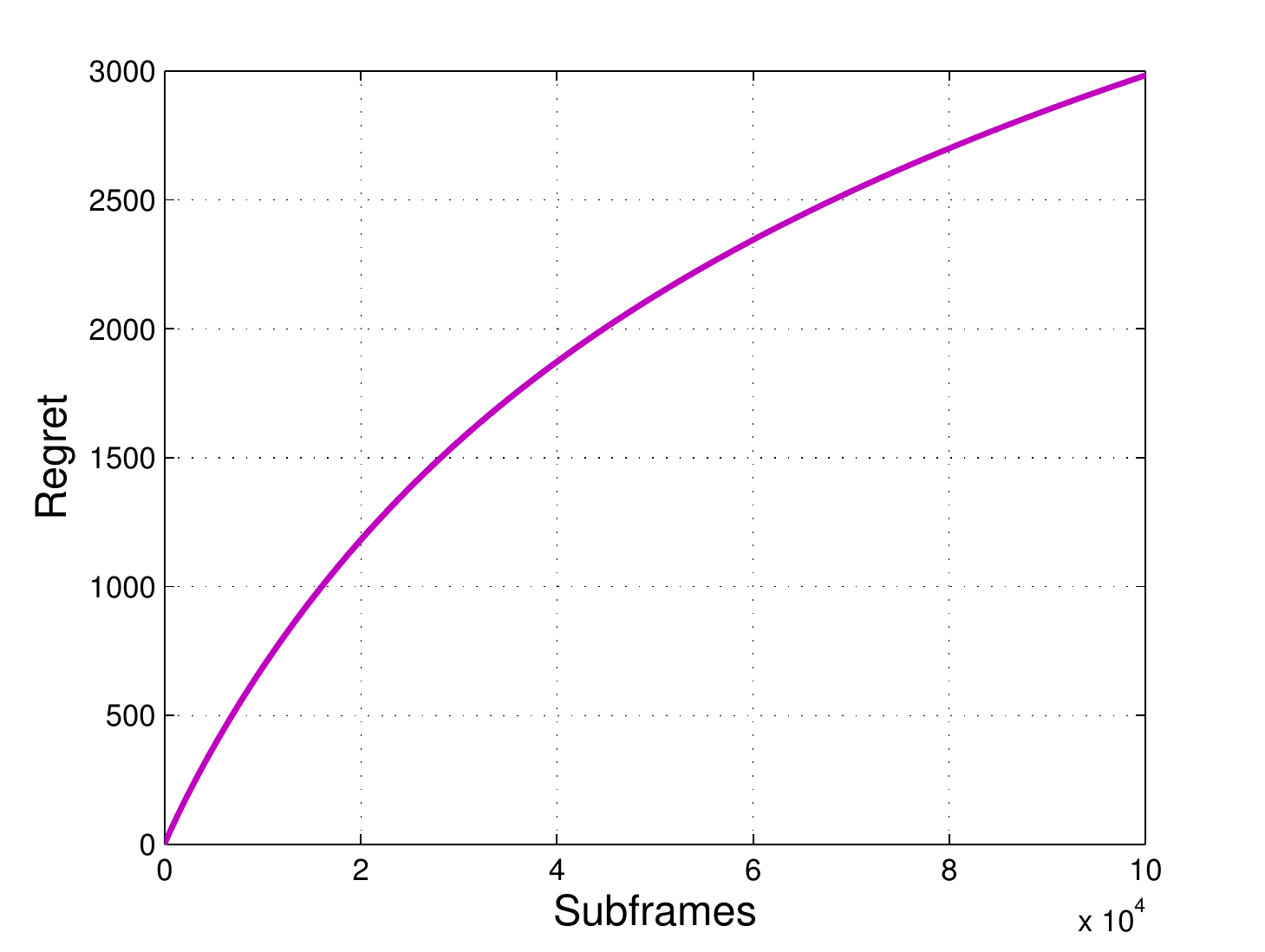}
                   \end{center}
                  \caption{Regret over subframes with UCB1 for a single D2D player.}\label{fig:2}
                  \end{figure} 
                                      \begin{figure}[t]
                                    \begin{center}
                                  \includegraphics[scale=0.56]{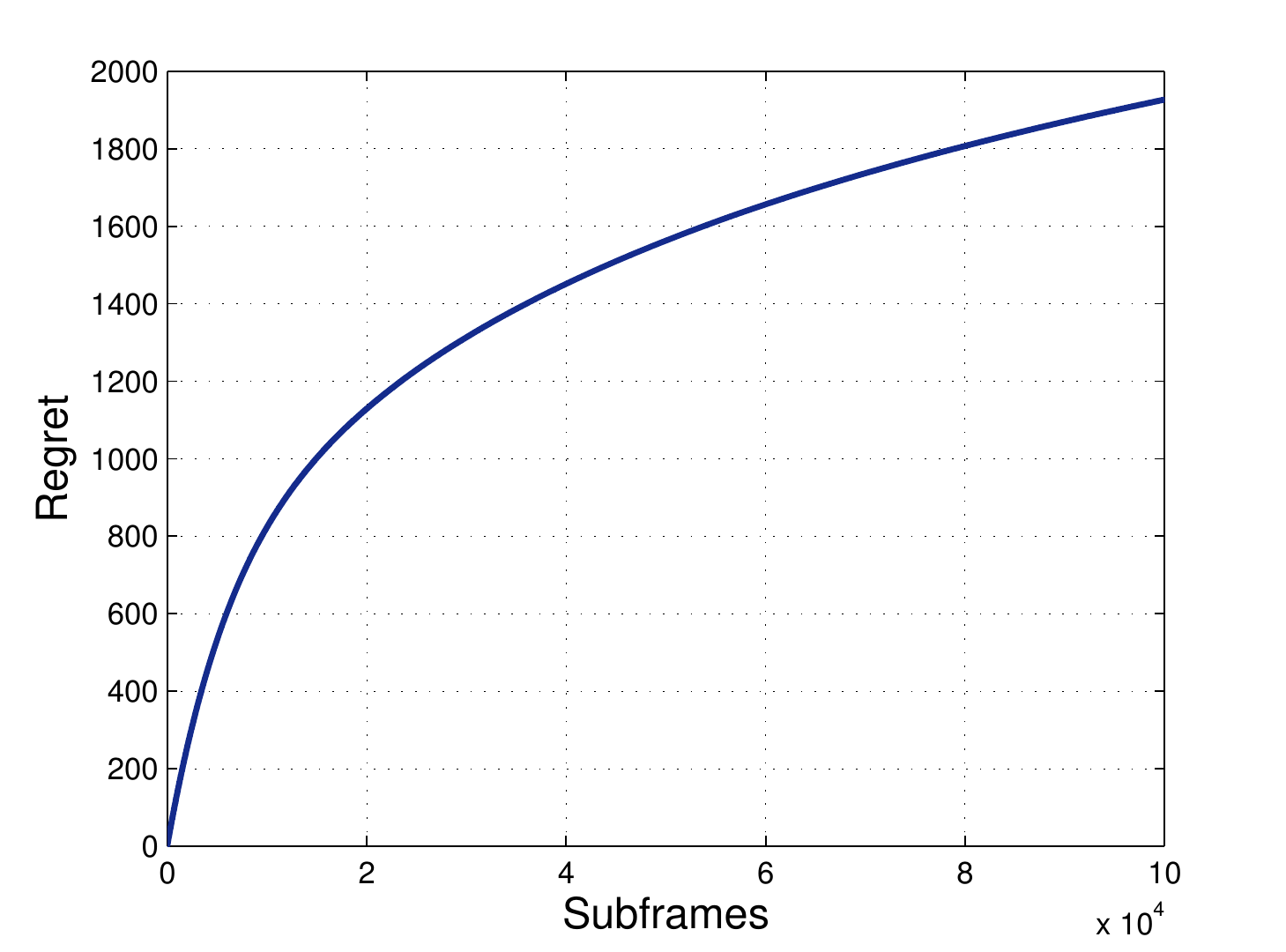}
                                   \end{center}
                                  \caption{Regret over subframes with Exp3 for a single D2D player.}\label{fig:3}
                                  \end{figure}      
 We model its gains by exponentially distributed random variables of mean 1. The transmit power  $P_c$ of a CU is set to 250 mW. The UE and the BS noise figures are 9 dB and 5 dB respectively. The
thermal noise density is set to -174 dBm. The SINR threshold $\gamma^{tgt}$ for the CUs is 10 dB. The total number of subframes is $10^5$, each of duration 1 ms. The transmission bandwidth is 180 kHz. For the $k$th-UCB1 based policy we set $\beta=50$. For the Exp3 based policy, we set $\alpha=0.01$ and $r'= 64$ kbps. Every plot is obtained after 50 Monte Carlo (MC) simulations over the policy with a fixed topology and 10 MC simulations over topologies. We next compare the performances of our proposed algorithms. \\ 
									\begin{figure}[t]
                                   \begin{center}
                                 \includegraphics[scale=0.56]{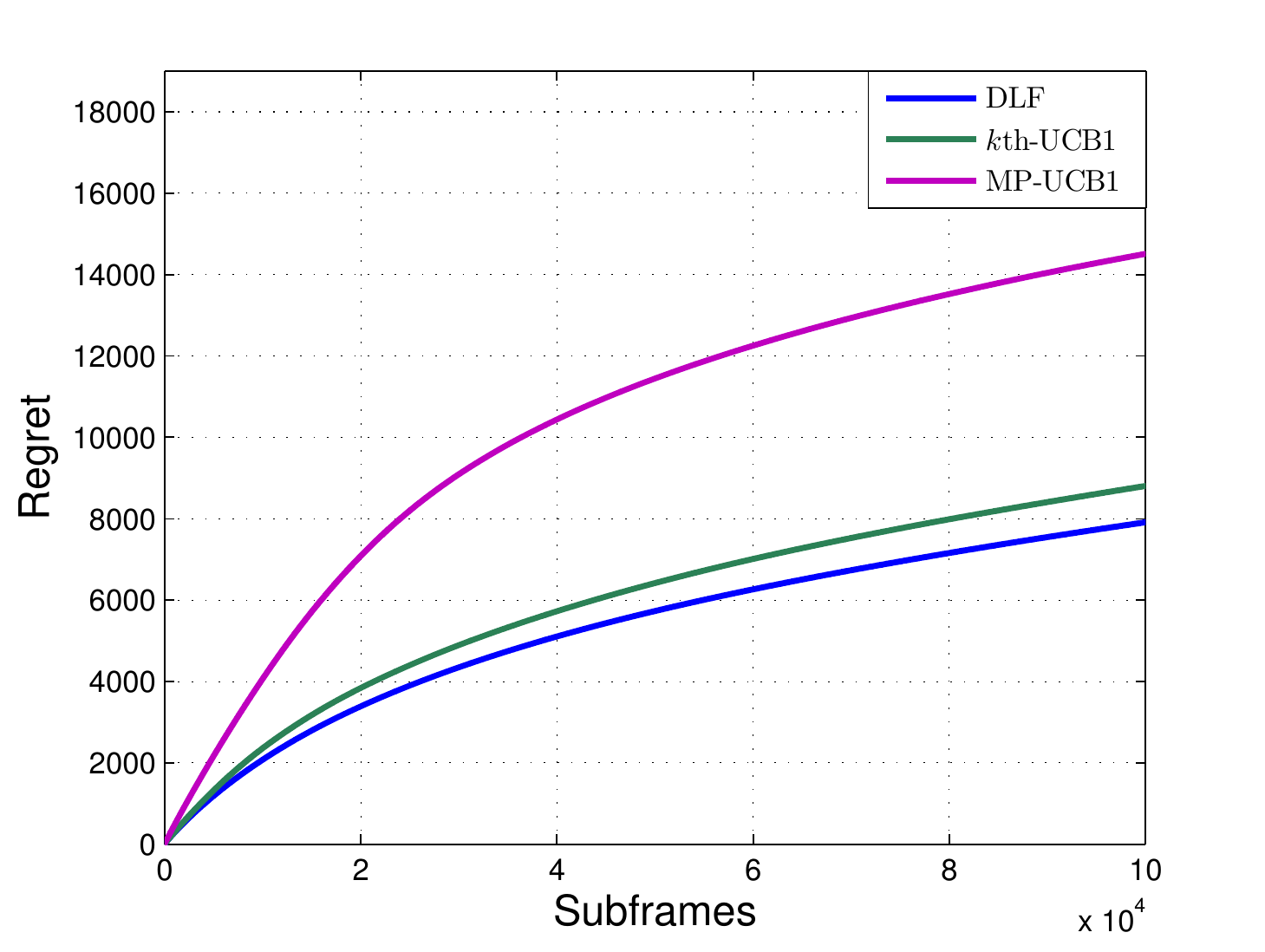}
                                  \end{center}
                                 \caption{Regret over subframes with MP-UCB1, DLF and $k$th-UCB1 for multiple D2D players.}\label{fig:4}
                                 \end{figure}  
                                     \begin{figure}[t]
                                   \begin{center}
                                 \includegraphics[scale=0.56]{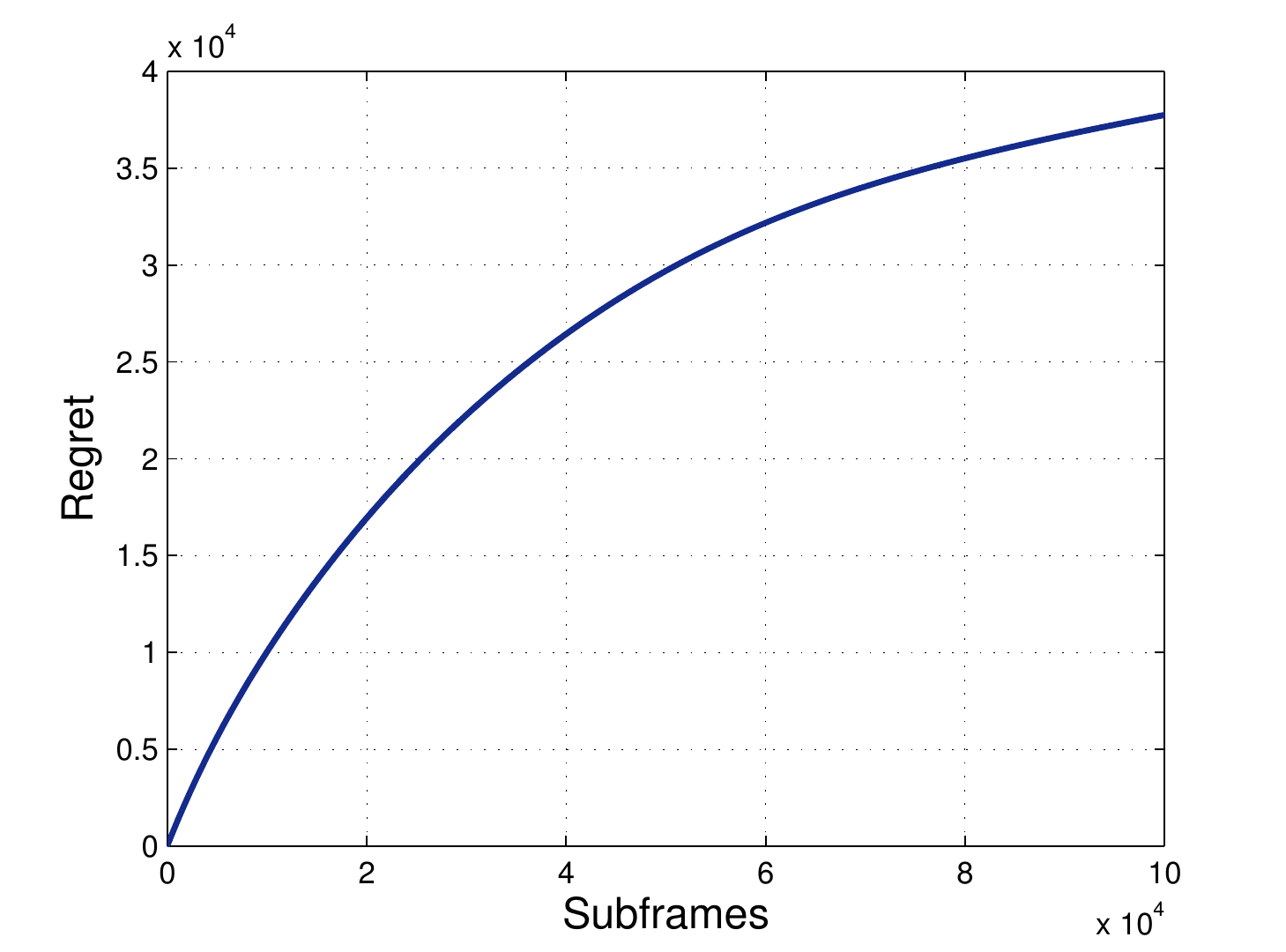}
                                  \end{center}
                                 \caption{Regret over subframes with Exp3 for multiple D2D players.}\label{fig:5}
                                 \end{figure}                                 
\noindent \emph{Regret:} For the MP-UCB1,  DLF and $k$th-UCB1 based policies, the distribution and the mean of the reward process of each arm are unknown. However, for the Exp3 based policy we model the reward in such a way that the distribution of each arm is Bernoulli but the mean of each arm is unknown. 
In both these cases the mean reward of an arm is different because the reward model is different for both of them. 

Thus, we cannot compare the regret of the Exp3 based policy with the regret of the index based policies even though the channel parameters are the same.    
For the index based allocation policies also we cannot compare the regrets of the MP-UCB1 based policy with DLF and $k$th-UCB1 based policies because the regret definition is different for them. For a single D2D player, the regret plots of our proposed policies with UCB1 and Exp3 are shown in Figs. \ref{fig:2} and \ref{fig:3} while for multiple D2D players, Fig. \ref{fig:4} demonstrates the regret plot of the index based policies and Fig. \ref{fig:5} demonstrates the regret plot of the Exp3 based policy. The regret of the $k$th-UCB1 based policy is more than DLF because it selects the arm with the $K^{th}$ largest UCB1 index with probability $1-\epsilon_n$ and explores the arms in $\mathcal{O}_K$ with probability $\epsilon_n$ in every subframe but the DLF based policy chooses an arm according to the metric of Eqn. 13 in every subframe and doesn't explore. \\
                                    \begin{figure}[t]
                                  \begin{center}
                                \includegraphics[scale=0.56]{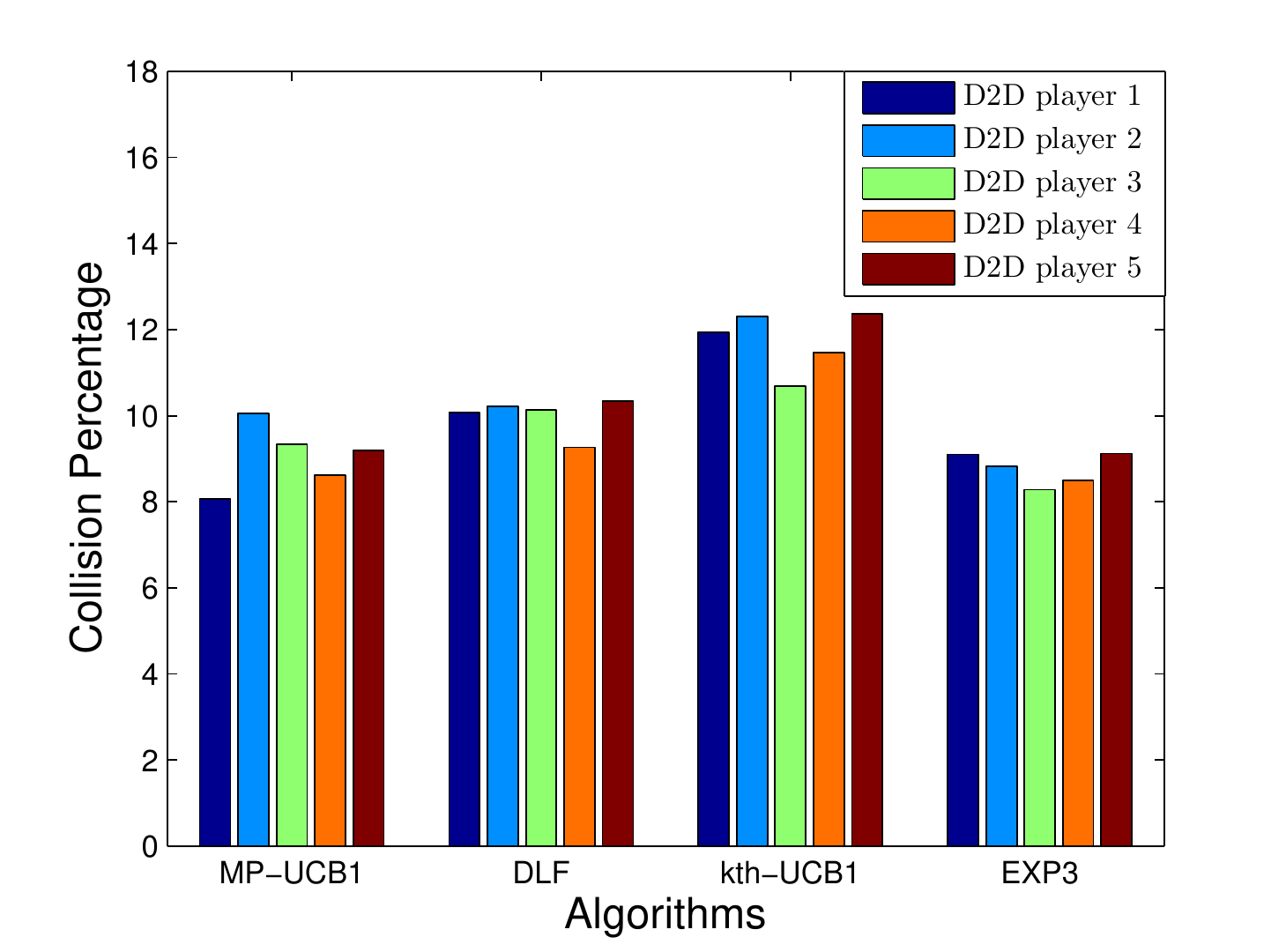}
                                 \end{center}
                                \caption{Bar graph showing collision percentage for each D2D player with MP-UCB1, DLF, $k$th-UCB1 and Exp3.}\label{fig:6}
                                \end{figure} 
                                         \begin{figure}[t]
                                       \begin{center}
                                     \includegraphics[scale=0.56]{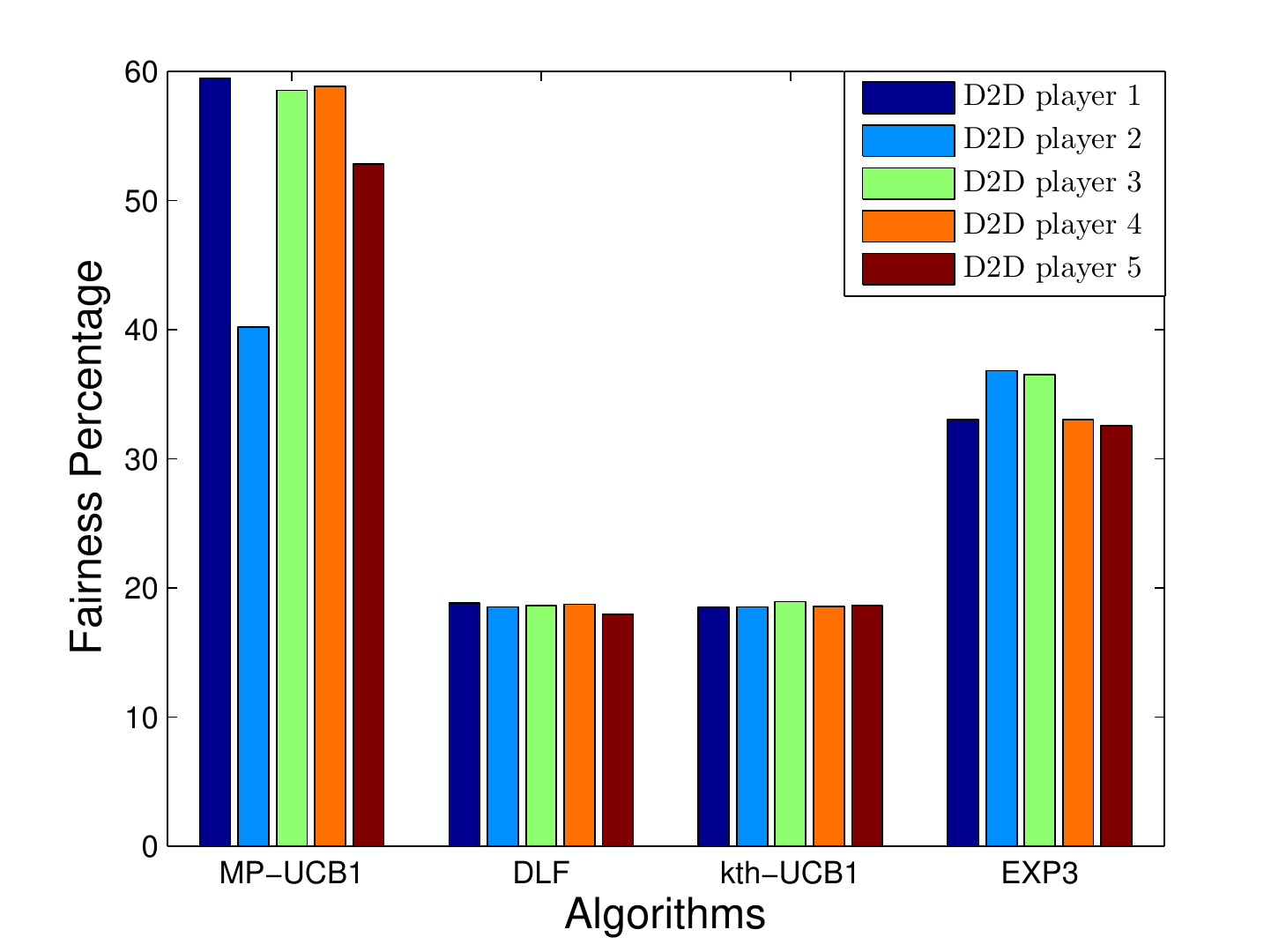}
                                      \end{center}
                                     \caption{Bar graph showing fairness of the algorithms}\label{fig:7}
                                     \end{figure} 
   \emph{Collision Percentage:} We observe from the bar graph of Fig. \ref{fig:6} that the D2D players collide the most for the $k$th-UCB1 based policy, followed by DLF. For these policies we had expected that the chances of two differently ranked D2D users selecting the same CU would reduce which would decrease the collisions. However, it is not so. We observe that the percentage collision is lesser for the MP-UCB1 and Exp3 based policies rather than $k$th-UCB1 and DLF.                          
   \begin{definition}
   We define an algorithm to be fair when each D2D player $d$ gets an equal opportunity over time of being allocated a CU that gives it the highest expected reward (index based algorithms) or the highest return (Exp3) over subframes. It is measured as the percentage of the number of times each D2D player is allocated this CU over the subframes. 
   \end{definition}
                                                      \begin{figure}[t]
                                                    \begin{center}
                                                  \includegraphics[scale=0.56]{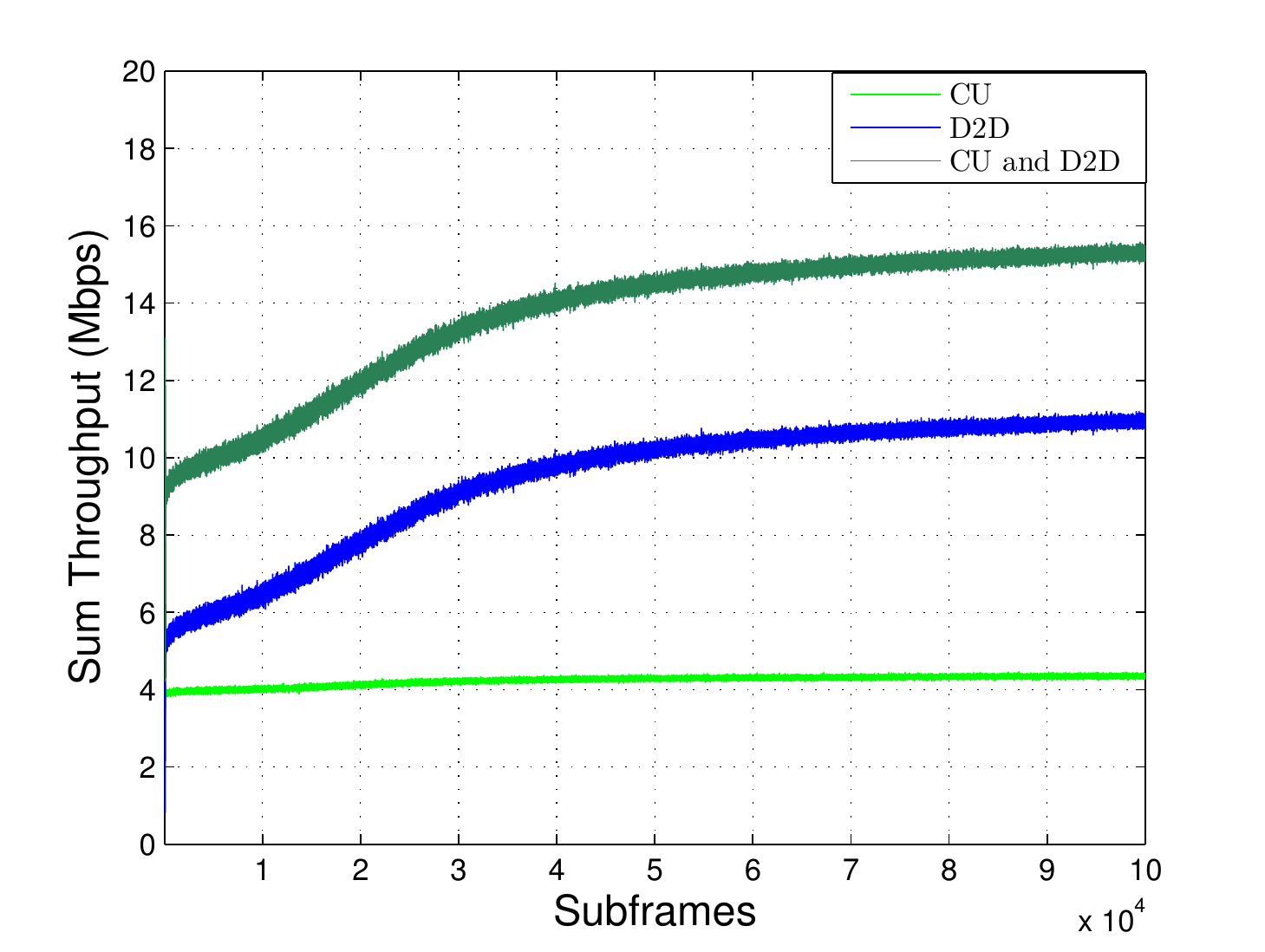}
                                                   \end{center}
                                                  \caption{ Sum throughput of the D2D players and the CUs with MP-UCB1.}  \label{fig:8}  
                                                  \end{figure}  
                                                          \begin{figure}[t]
                                                        \begin{center}
                                                      \includegraphics[scale=0.56]{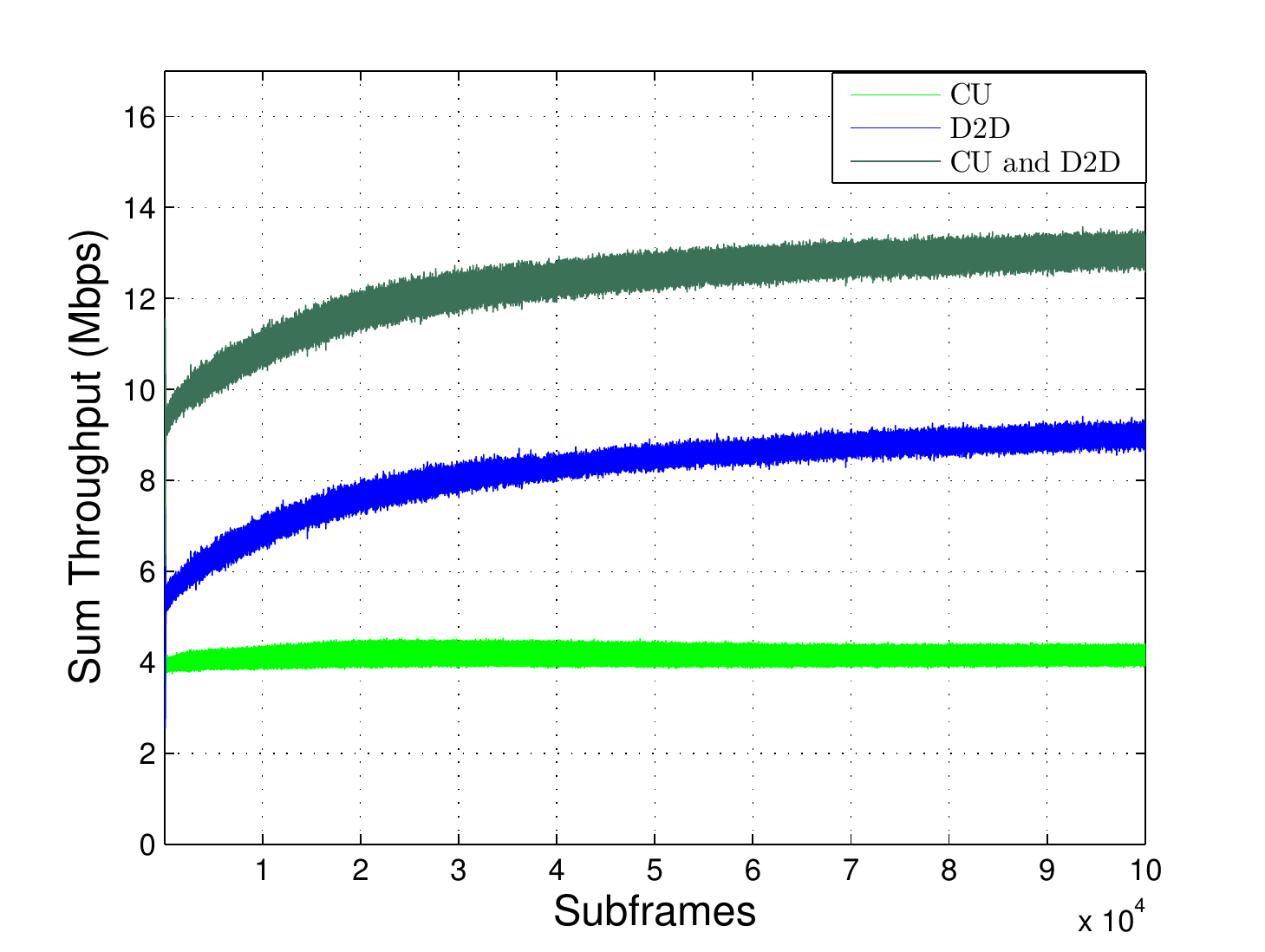}
                                                       \end{center}
                                                      \caption{Sum throughput of the D2D players and the CUs with DLF.}\label{fig:9}
                                                      \end{figure}   
   \noindent \emph{Fairness:} We observe from Fig. \ref{fig:7} that for all the algorithms every D2D player gets a fair chance to select the arm that gives it the highest expected reward in case of the index based policies or the highest return in case of Exp3 over the subframes. Note that for MP-UCB1, the percentage of selecting the arm with the highest expected reward for each D2D player is more as compared to $k$th-UCB1 and DLF because $k$th-UCB1 and DLF rank the players and MP-UCB1 doesn't. Each D2D player in MP-UCB1 (or Exp3) tries to choose the CU that gives it the highest expected reward (or highest returns) in every subframe but for $k$th-UCB1 and DLF each D2D player tries to choose the CU that gives it the highest expected reward after every $N_D-1$ subframes when its turn comes.\\

\begin{figure}[t]
                                            \begin{center}
                                          \includegraphics[scale=0.56]{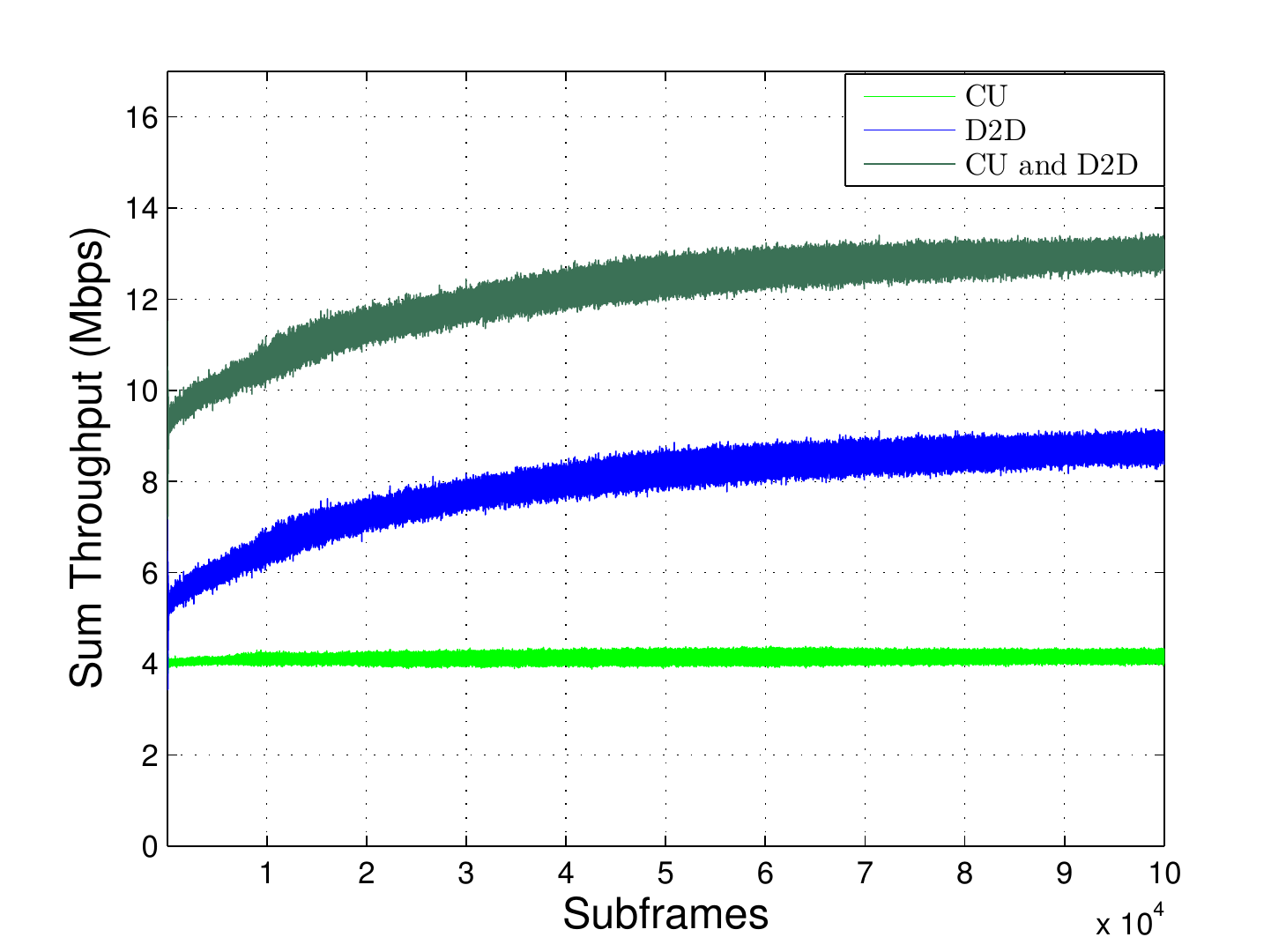}
                                           \end{center}
                                          \caption{Sum throughput of the D2D players and the CUs with $k$th-UCB1.}\label{fig:10}
                                          \end{figure}

\begin{figure}[t]
                                            \begin{center}
                                          \includegraphics[scale=0.56]{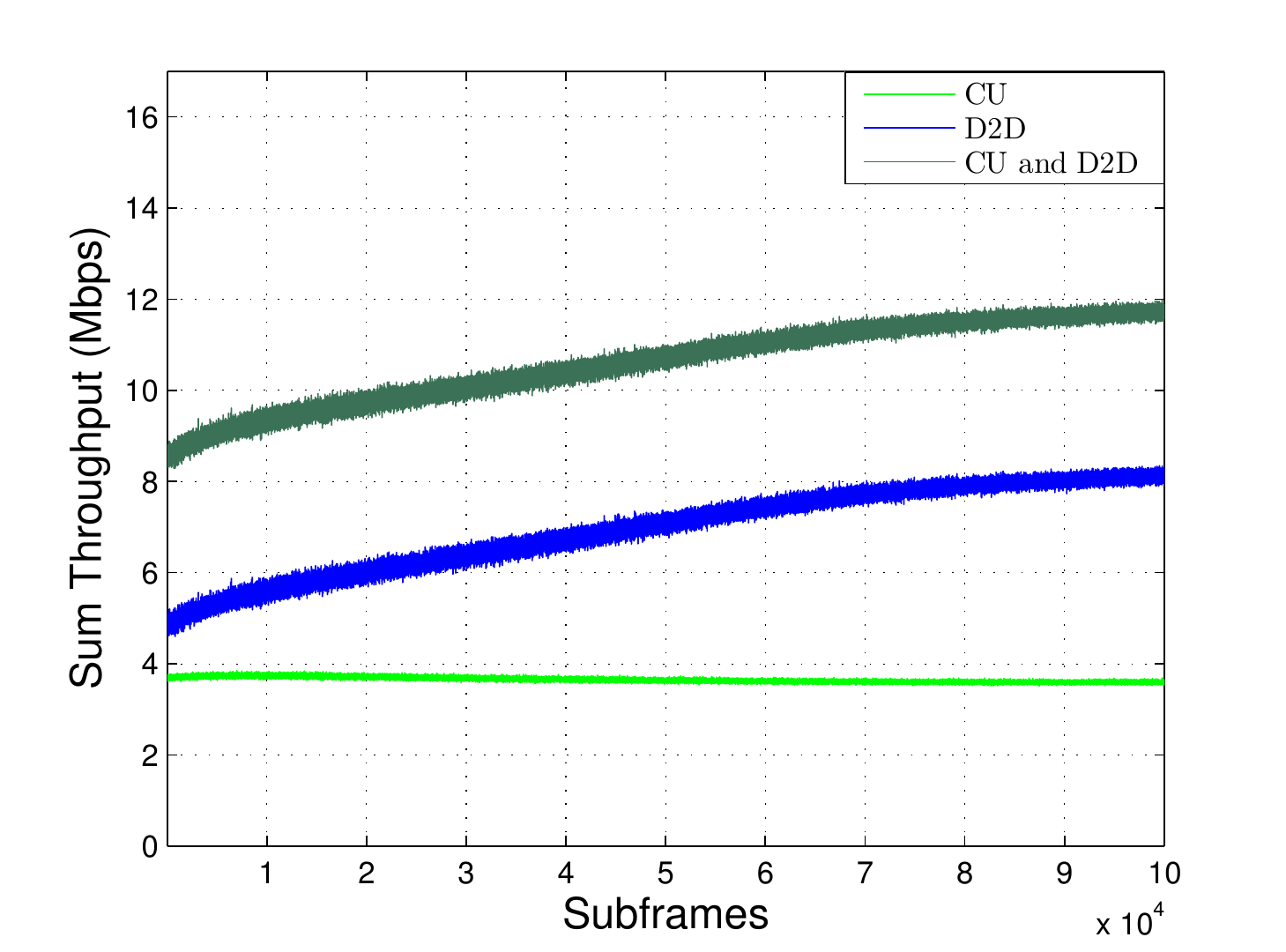}
                                           \end{center}
                                          \caption{Sum throughput of the D2D players and the CUs with Exp3.}\label{fig:11}
                                          \end{figure}

\emph{Sum Throughput:}
    1) We observe that the sum throughput of the D2D players is less initially as they suffer from more collisions due to which the power allocated to them becomes zero. 2) The plots of Fig. \ref{fig:8} - \ref{fig:11} depict that the sum throughput of the D2D players in the long run is higher for the MP-UCB1 based policy as compared to $k$th-UCB1, DLF and Exp3 based policies. For MP-UCB1, the D2D players collide less as seen from Fig. 5.6 and percentage of the number of times that they try to select their optimal CUs (fairness percentage) is also high as compared to the other algorithms. Thus, the sum throughput of the D2D players is higher.  For Exp3 the fairness percentage is less which means that they are not able to select their optimal CUs over subframes. Because of this, the sum throughput of the D2D players is less in the long run. For $k$th-UCB1 and DLF though the collision percentage is high and the fairness percentage is low, the sum throughput of the D2D players is still high because each D2D player is successful in trying to choose the arm which gives it the $K^{th}$ largest expected rewards as per its rank $K$. 
However, the sum throughput of the D2D players for all the four algorithms are comparable.
   3) We have plotted the sum throughput of the 5 CUs that the D2D players select out of 20 CUs in every subframe in order to reuse their resources. We observe from the plots that the sum throughput of the CUs are maintained at a constant rate on an average. The fluctuations of the sum throughput of the CUs are because of the following reasons. When a D2D player selects a CU,  if the SNR of the CU is less than $\gamma^{tgt}$ then power is not allocated to it otherwise its SINR would become lower than $\gamma^{tgt}$. This implies that the CU's throughput will be lower than $r^{tgt}$. Moreover, if the D2D player collides with another D2D player, then also it is not allocated any power and hence the CU's throughput can be of any value. However, when it is allocated power, the CU's throughput gets fixed at $r^{tgt}$.  When the power allocated to the D2D player is limited to $P_{max}$, when more power could have been allocated to it theoretically, the CU's rate becomes greater than $r^{tgt}$. 4) We observe that the sum throughput of the D2D players is more than that of the CUs. This is so because the communication range of the D2D transmitter and receiver is small as compared to the CU and the BS. 
\section{Conclusions}
In this work, we consider the problem of power and resource allocation to the D2D players with partial CSI. We formulate this problem within the framework of MP-MAB. Our proposed policies ensure that QoS is guaranteed to the CUs by using the excess SNR of the CUs above a certain threshold to allocate power to the D2D players. We propose two optimal learning policies based on UCB1 and Exp3 for a single player 
and then extend them to a multi-player setting. By applying the Exp3 algorithm we show that the D2D resource allocation problem can also be solved within an adversarial MP-MAB framework.  We propose two more policies based on DLF and $k$th-UCB1.  
We demonstrate through the simulation results that our proposed policies are fair  and perform well. By comparing them, we found that MP-UCB1 performs better than the others for this problem.

\end{document}